\documentclass{article}
\usepackage{RR}

\usepackage[english]{babel}

\usepackage{ifpdf}
\usepackage{amsmath}
\usepackage{amssymb}
\usepackage{amsfonts}
\usepackage{amsthm}
\usepackage{subfigure}
\usepackage{mathrsfs}
\usepackage[mathcal]{euscript}
\newtheorem{theorem}{Theorem}
\newtheorem{lemma}[theorem]{Lemma}
\newtheorem{corollary}[theorem]{Corollary}

\renewcommand\geq\geqslant
\renewcommand\leq\leqslant
\newcommand\OO{\mathscr{O}}

\ifpdf
  \usepackage{aeguill}
\else
\fi

\RRdate{Février 2007}

\RRetitle{An Upper Bound on the Average Size of Silhouettes}
\RRtitle{Une borne supérieure sur la taille moyenne des silhouettes}
\RRauthor{Marc Glisse\thanks{INRIA Lorraine - Universit{\'e} Nancy~2, LORIA, Campus Scientifique, B.P. 239,
54506 Nancy, France. {\tt Firstname.Name@loria.fr}.}
\and
Sylvain Lazard\footnotemark[1]
}
\begin{document}
%\conferenceinfo{SCG'06,} {June 5--7, 2006, Sedona, Arizona, USA.}
%\CopyrightYear{2006}
%\crdata{1-59593-340-9/06/0006}

%\section*{Abstract}
\RRabstract{%
It is a widely observed phenomenon in computer graphics that the size of
the silhouette of a polyhedron is much smaller than the size of the whole
polyhedron. This paper provides, for the first time, theoretical evidence
supporting this for a large class of objects, namely for polyhedra that
approximate surfaces in some reasonable way; the surfaces may be
non-convex and non-differentiable and they may have boundaries. We prove
that such polyhedra have silhouettes of expected size $O(\sqrt{n})$
where the average is taken over all points of view and n is the
complexity of the polyhedron.}

\RRresume{%
Il est connu en infographie que la taille de la silhouette d'un polyèdre
s'avère souvent, en pratique, bien plus petite que la taille du polyèdre
entier. Cet article est le premier à fournir des preuves théoriques
justifiant cette observation pour une large classe d'objets~: les
polyèdres qui approximent des surfaces de manière raisonnable~; les surfaces
considérées ne sont pas nécessairement convexes ou lisses et elles
peuvent avoir un bord. Nous prouvons qu'un tel polyèdre de taille n a, en
moyennant sur tous les points de vue à l'infini, une silhouette de taille
$O(\sqrt{n})$. }

%\category{F.2.2}{Analysis of Algorithms and Problem Complexity}{Nonnumerical Algorithms and Problems}[Geometrical problems and computations]
%\terms{Algorithms, Theory}
\RRkeyword{silhouette, polyhedron, upper bound}
\RRmotcle{silhouette, polyèdre, borne supérieure}

\RRprojet{Végas}

\RRtheme{\THSym}

\URLorraine

\makeRR

%It is a widely observed phenomenon in computer graphics that the size of
%the silhouette of a polyhedron is much smaller than the size of the whole
%polyhedron. This paper provides for the first time theoretical evidence
%supporting this. We prove that for a large class of objects and a random
%point of view, the silhouette of a polyhedron of size \(n\) has size
%\(O(\sqrt{n})\).

%This paper provides for the first time theoretical evidence supporting 
%the widely observed phenomenon that the silhouette of a large class of
%objects has expected size \(O(\sqrt{n})\).

\section{Introduction}
The silhouette of a polyhedron with respect to a given viewpoint is,
roughly speaking, the set of
edges incident to a front and a back face.
%see Figure~\ref{fig=silex} for an illustration.
Silhouettes arise in various
problems in computer graphics such as hidden surface removal
and shadow computations (see
\cite{d-scurev-04,dd-rev-02,eghz-irsmps-00} for some recent references)
and algorithms to compute them efficiently have been well-studied (see
the survey by Isenberg et al. \cite{ifhss-dgsap-03}).
They are important in shape recognition;
Sander et al. \cite{sgghs-sc-00} claim that the silhouette
``is one of the strongest visual cues of the shape of an object''.
%The computation of the silhouette has been well-studied.
%A survey of
%techniques
%for the fast computation of polyhedral silhouettes
%can be found in Isenberg et al. \cite{ifhss-dgsap-03}.

It is a widely accepted fact that the silhouette of a polyhedron is
usually much smaller than the whole polyhedron. Sander et al.
\cite{sgghs-sc-00}, for instance, state the largely repeated claim
that the silhouette of a mesh is
often of size \(\Theta(\sqrt{n})\) where \(n\) is the number of faces of the
mesh. An experimental study by Kettner and
Welzl \cite{kw-ceafpp-97} confirms this for a set of realistic objects.
This experimental study was extended by McGuire \cite{m-oss-04} to a larger
database of larger objects
%, and the claim of that paper is that
for which the observed size of
the silhouette is approximately
%of size
\(n^{0.8}\).

There are few theoretical results supporting these observations.
Kettner and
Welzl \cite{kw-ceafpp-97} prove that a convex polyhedron that approximates a
sphere with Hausdorff distance \(\varepsilon\) has
\(\Theta(1/\varepsilon)\) edges, and a random orthographic projection of
such a polytope has \(\Theta(1/\sqrt{\varepsilon})\) silhouette edges.
Alt et al. \cite{agg-owcc-03} give conditions under which it can be
proved that the average silhouette of a \emph{convex} polyhedron has size
\(O(\sqrt{n})\) and give additional conditions under which the worst-case
size is provably sub-linear.

\begin{figure}[b]
\centering
\subfigure{%
\includegraphics[height=4.5cm]{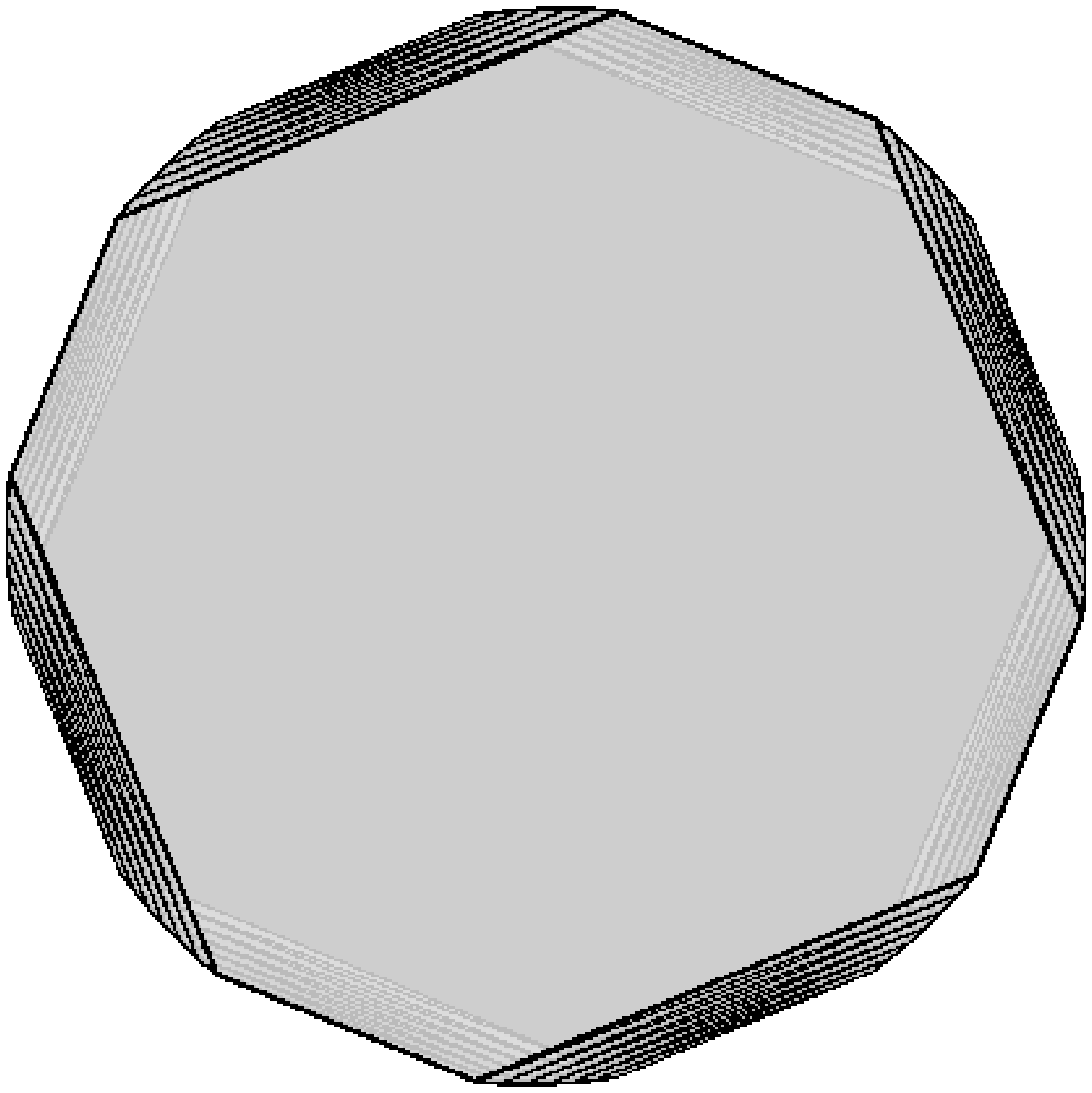}%
}
\subfigure{%
\includegraphics[height=4.2cm]{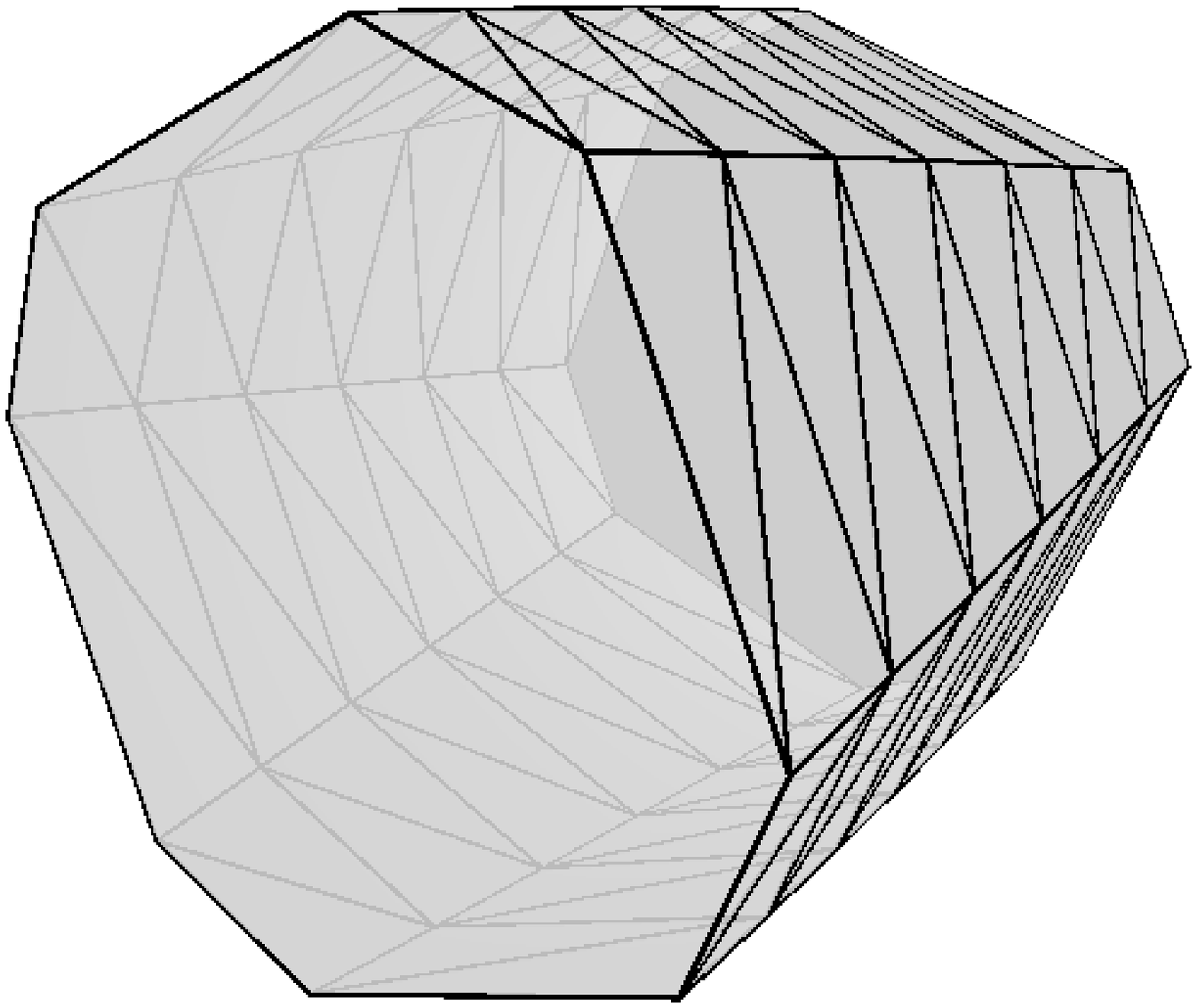}%
}
\label{fig=cyl}
\caption{A worst-case linear silhouette (left) of a
polyhedron approximating a cylinder.}
\end{figure}

The goal of this paper is to study the average silhouette size of
\emph{non-convex} polyhedra. Convexity is a very strong assumption, which was
crucial in the previous theoretical results. Here, rather, we assume that the
polyhedron is a good approximation of some fixed (not necessarily convex)
surface. Notice that it is very difficult to guarantee anything on the {\it
  worst-case} complexity of the silhouette of a polyhedron unless it
approximates a strictly convex surface. Alt et al. \cite{agg-owcc-03} give an
example of a polyhedral approximation of a section of a cylinder with worst-case silhouette
size \(\Theta(n)\) (see Figure~\ref{fig=cyl}). Moreover, their example can be
modified in such a way that the surface is smooth, and its polyhedral
approximation is as ``nice'' as one might hope (for instance, it can be required
that the faces are fat triangles that all have almost the same size).

In this paper we prove an upper bound on the {\it expected size} of the
silhouette for random viewpoints.
We prove that the silhouette of a polyhedron that approximates a surface
in a reasonable way has expected size \(O(\sqrt{n})\). Note that the
average is taken over all viewpoints for a given surface, and
not on a set of surfaces.

In Section~\ref{sec=def}, we define precisely the notion of silhouette
for polyhedra and general surfaces. We then present and prove our main
result in Section~\ref{sec=result} and conclude in Section~\ref{sec=concl}.

\section{Definitions}
\label{sec=def}

\begin{figure}
\centering
\subfigure[]{%
\includegraphics[height=4cm]{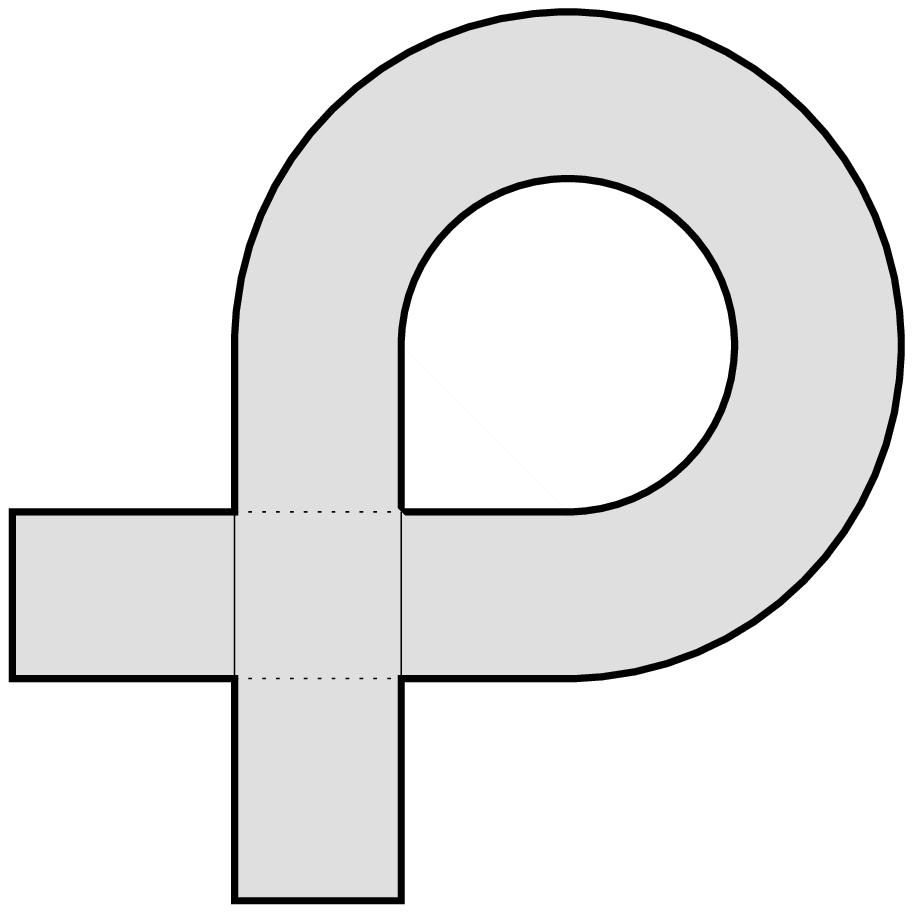}%
\label{fig=silex1}%
}
\subfigure[]{%
\includegraphics[height=4cm]{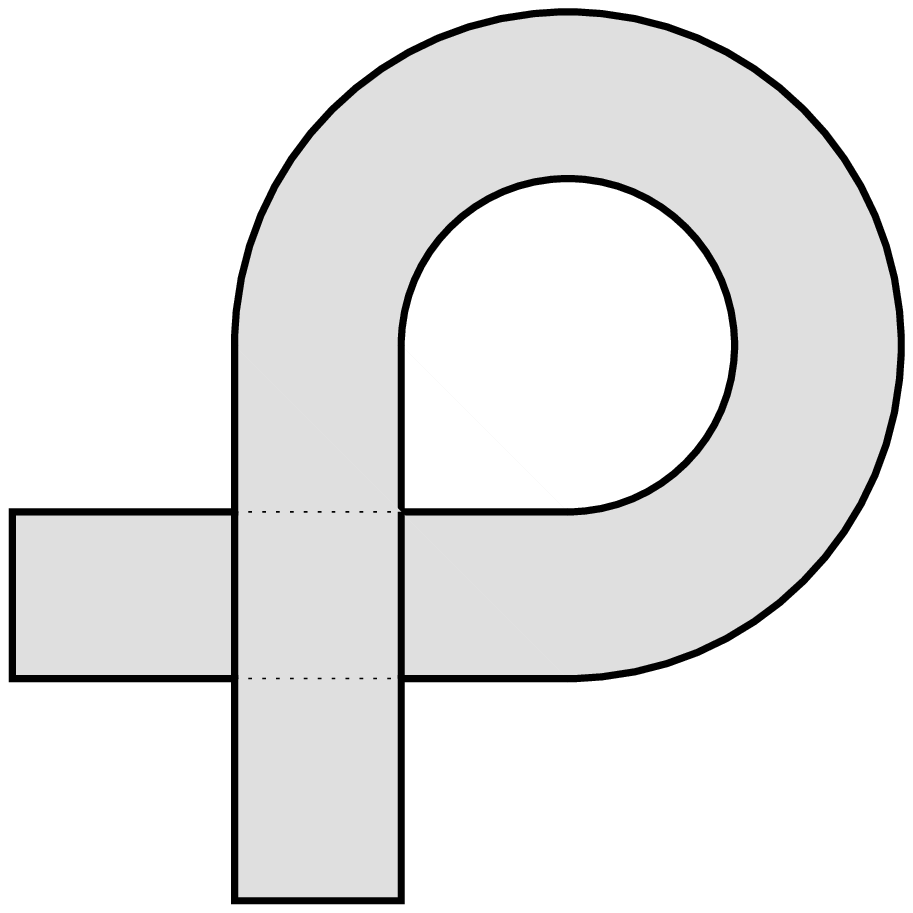}%
\label{fig=silex2}%
}
\subfigure[]{%
\includegraphics[height=4cm]{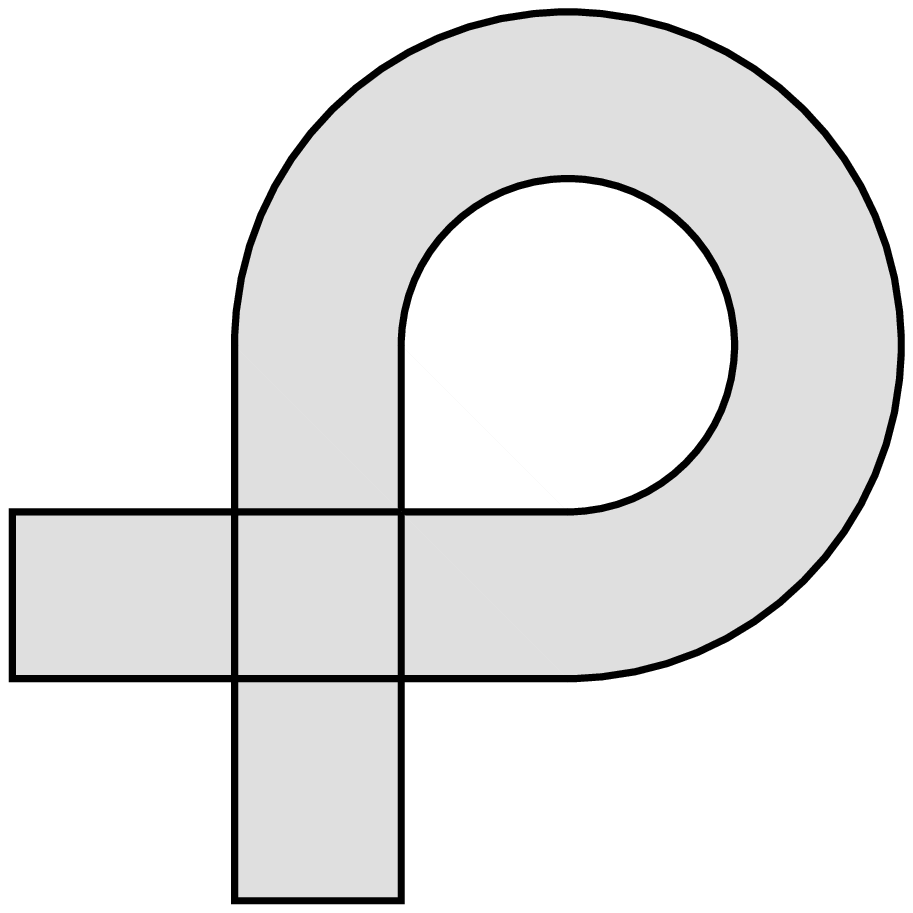}%
\label{fig=silex3}%
}
%\caption{illustration of silhouettes including, or not, self-cast
%shadows and considering semi-transparent objects.}
%-%\caption{Three different notions of silhouettes: (a) contour, (b) apparent
%-%boundary, and (c) transparent silhouette.}
\caption{Three different notions of silhouettes: (a) outline of a solid,
as cast by its shadow, (b) rim of an opaque object, and (c) rim of a
transparent object.}
\label{fig=silex}
\end{figure}

The term silhouette has been used in the literature to represent several
different notions, depending on the application, reflecting such issues
as: is the object considered opaque or transparent? Is occlusion taken
into account? Is one interested by what the eye perceives, {\it i.e.,} a plane
curve, or by the space curve which gave birth to it? In the area of
photography, for instance, a silhouette (also called apparent boundary)
is defined as an outline of a
solid object, as cast by its shadow, that appears dark against a light
background (Figure~\ref{fig=silex1}). In the field of computer vision, by
contrast, the silhouette (also called rim, profile or contour generator) is
roughly defined as the curve on the surface that separates front face
regions from the back ones, either for opaque (Figure~\ref{fig=silex2})
or for transparent (Figure~\ref{fig=silex3}) objects.

In this paper we prove an upper bound on the size of the transparent
silhouette; since such a silhouette contains the apparent boundary and the
contour, our bounds also apply to all these types of silhouettes. In the rest
of the paper the term silhouette will be used to mean transparent
silhouette.

In the rest of this section we give a formal definition of silhouettes of
polyhedra and then provide a definition for more general surfaces.

%%
%%There are several possible definitions for the silhouette of a surface.
%%Basically, the silhouette can be defined as the shadow cast by a point
%%light source and an opaque object on a wall (see
%%Figure~\ref{fig=silex1}). It may also include the shadows the object
%%casts on itself (see Figure~\ref{fig=silex2}). The silhouette can also be
%%defined by considering the object as semi-transparent (see
%%Figure~\ref{fig=silex3}). Note that these silhouettes can be defined
%%equivalently as the contour of the object observed from a viewpoint.
%%
%%We consider in this paper only the type of silhouette where the objects
%%are semi-transparent. Note that these
%%(semi-transparent) silhouettes contain the other two types of
%%silhouettes. Our results thus apply to all three types of silhouettes.
%%
%%We first give a formal (though intuitive) definition of silhouettes of
%%polyhedra. We then provide a definition for more general surfaces.

\subsection{Polyhedra}

%\begin{itemize}
%\item

The (transparent) \emph{silhouette} of a polyhedron from a viewpoint
(possibly at infinity) is the set of edges
that are adjacent to a front face and a back face. A face is considered a
front face if the angle between its normal vector and a vector from a
point of the face to the viewpoint is acute, and a back face if that
angle is larger than \(\pi/2\). If the point of view is
in the plane containing the face,
we refer to the definition of silhouettes for the case of general
surfaces.
%the face is considered\footnote{%
%In order to get a definition that is consistent with the one for
%general surfaces, we need to consider, for each point on the edge, the
%other edge that the line passing through it and the viewpoint intersects
%in this face and check the orientation of the next face. But since this
%happens for a negligible set of viewpoints, we can forget it here.%
%}
%a back face
%(considering all these faces as front faces would yield an equally
%coherent definition).
The normal vectors should point outwards, but what really matters is that
the orientation is consistent for the two faces that share this edge, so
this definition also applies to non-orientable (necessarily
self-intersecting) polyhedra.

In this paper, we call complexity of a silhouette (of a polyhedron)
its number of edges.

%%\item 
%
%The \emph{complete silhouette} consists of the (parts of) edges of the
%transparent silhouette that are visible from \(C\) (where a point \(A\)
%is said to be visible is there are points \((A_n)_{n\in\mathbb{N}}\)
%arbitrarily close to \(A\) such that the segments \([CA_n]\) do not
%intersect the surface).
%
%%\item
%
%The \emph{apparent boundary} consists of the (parts of) edges of the
%transparent silhouette in the neighborhood of which \(C\) can see at
%infinity.

%\end{itemize}

\subsection{General surfaces}

Our objective is to bound  the size of the
silhouette of a polyhedron. To achieve this goal, we need to relate
the silhouette of the polyhedron to the silhouette of the surface it 
approximates, which means we need a definition of silhouettes that
applies to a larger class of objects. Although this may seem unintuitive, we first define the silhouette as a set of rays, and
then relate this to the more usual
%points-on-the-surface view.
concept of a set of points on the surface.

Let \(S\) be a compact 2-manifold without boundary. It separates
\(\mathbb{R}^3\) in two non-empty open regions; call \(\OO\) and \(\OO'\) their
closures (so \(\OO\cap \OO'=S\) and \(\OO\cup \OO'=\mathbb{R}^3\)).  Let $V$ be
a viewpoint not on \(S\) but possibly at infinity.  The (transparent)
\emph{silhouette} of \(S\) from \(V\) is the set of rays \(R\) starting from
\(V\) that are tangent to
\(S\) in a non-crossing way (\(R\) may cross \(S\) elsewhere). More
formally, we require that there exists an open segment \(u\) of \(R\) that
contains a connected component of
\(R\cap S\) and is contained either in \(\OO\) or \(\OO'\).%
%
%\footnote{%
%One can give an alternative definition: \(R\) is tangent to \(S\) if and
%only if there exists an open neighborhood \(U\) in \(S\) of a connected
%component of \(R\cap S\) such that in any neighborhood of \(R\) there
%exists a ray that does not intersect \(U\). This may define a slightly
%different set of tangents in complicated cases. It also has a
%generalization to immersions.%
%}%

This definition defines a set of rays. The silhouette can also be seen
as the trace of this set of rays on the surface.
%More precisely, for a
%ray \(R\) on the silhouette, we add to the silhouette the
%union of the connected components of \(R\cap S\) that satisfy the
%non-crossing property.
%
%Alternatively, for a ray \(R\) and a set of points on the silhouette
%as previously defined, we can keep only
%one point for each connected component, namely the one closest to \(V\).
More precisely, for each ray \(R\) on the silhouette, we consider the
closest point to \(V\) on each connected component of \(R\cap S\) that
satisfies the non-crossing property.
This definition is consistent with the one given for the particular case
of polyhedra, and is the one we will use in this paper.

For a given viewpoint at infinity, we define the (projected) \emph{length} of
the silhouette as the length (counted with multiplicity if several points
have the same projection) of the projection of the silhouette, along the
direction given by the viewpoint, on an orthogonal plane.

\medskip
%{\flushleft \subsecfnt Remark: \ }
%TODO: virer le paragraphe suivant
\noindent{\bf Remark.}
%\footnote{%
The definition of the silhouette can be extended to cases where \(S\) is
not a 2-manifold, but an immersion of a compact 2-manifold. More
precisely, we have a 2-manifold \(S'\) and an application
\(f:S'\rightarrow \mathbb{R}^3\) such that \(S=f(S')\) and for any point
on \(S'\) there exists a neighborhood \(U\) of that point such that
\(U\) and \(f(U)\) are homeomorphic.
The local orientation is sufficient to decide whether \(R\) crosses \(S\)
or not (note that more complicated things can happen than crossing or
being tangent, even with smooth surfaces; for instance, the surface may ripple an
infinite number of times in the neighborhood of a point, making it
impossible to define on which side of \(S\) \(R\) is near the
intersection point).
This remark extends to the whole paper and, in particular,
to Theorem~\ref{th=principal}.
However, we do not give either a definition or a proof of this, as it
would uselessly make everything more obscure.%
%}%

\section{Main results}
\label{sec=result}

Let \(S\) be a compact 2-manifold without boundary whose silhouettes
have finite average length, $\mbox{silh}(S)$, where the average is taken over all
viewpoints at infinity. Let \(P_n\) be a
polyhedron with \(n\) triangular faces,
%TODO: discuss triangular
that is homeomorphic to \(S\) through \(f_n:P_n\to S\), such
that:

\begin{enumerate}

\item
\label{hyp-length}
the length of any edge of \(P_n\) is at least \(\frac{\alpha}{\sqrt{n}}\)
and

\item
\label{hyp-dist-bis}
for any point \(x\) on \(P_n\), \(d(x,f_n(x)) < \frac{\beta\, h(x)}
{\sqrt{n}}\) where \(h(x)\) is
%the length of an edge of a
the smallest height of the triangle(s) of $P_n$ that contain(s)~\(x\),

%\item
%\label{hyp-captain}
%the captain is at least 42 years old.

\end{enumerate}
where \(\alpha\) and \(\beta\) are two arbitrary positive numbers and
$d()$ denotes the Euclidean distance.

\begin{theorem}
\label{th=principal}
The expected complexity of the silhouette of
\(P_n\) is \(O\left(\sqrt{n}\right)\), where
the average is taken over all viewpoints at infinity. 
More precisely, for any $n$, the expected complexity is at most 
\[ \left(15\,\beta+\frac{24}{\alpha}\,\mbox{silh}(S)\right)\, \sqrt{n}.\]
%\[ \left(20\,\beta+\frac{24}{\alpha}\,\mbox{silh}(S)\right)\, \sqrt{n}.\]
\end{theorem}

Note that the bound is valid for any $n$ and any polyhedron $P_n$ satisfying the
above assumptions. Note also that the bound depends on $S$ only by the average
length of its silhouette.

We first clarify the meaning of the hypotheses on \(P_n\) and their
implications. We then prove Theorem~\ref{th=principal} in
Section~\ref{sec=proof}. We finally
show in Section~\ref{sec=general} how Theorem~\ref{th=principal} can
be generalized to surfaces with boundary and viewpoints at finite
distance.  In particular, we prove the following result.

Let $S'$ be  any compact  two-manifold with boundary of finite 
length  and whose silhouette has finite average length  (taken
over all viewpoints at infinity). 
\begin{theorem}
\label{th=principal2}
Any mesh $P_n$ with $n$ triangular faces that
approximates $S'$ according to Hypotheses 1 and 2 has a silhouette of expected
complexity $O(\sqrt{n})$ when the viewpoint is chosen uniformly at random in a
ball.
\end{theorem}

\subsection{Meaning of the hypotheses}
\label{sec=meaning}

Hypothesis~\ref{hyp-length} is here to avoid short edges. The main idea
of the proof is to link the complexity of the silhouette to its length,
and arbitrarily short edges would make this impossible. Now the
\(\frac{1}{\sqrt{n}}\) factor makes sense: intuitively, since the
polyhedron has \(n\) faces, each face has area of order \(\frac{1}{n}\), which
means that the edges have length of order \(\frac{1}{\sqrt{n}}\).

Hypothesis~\ref{hyp-dist-bis} is rather technical, and we discuss
instead the meaning of the following two more intuitive hypotheses,
which,
together with Hypothesis~\ref{hyp-length}, imply%
\footnote{%
Indeed, for any \(x\) in \(P_n\), Hypotheses~\ref{hyp-length}
and~\ref{hyp-fat} imply
that \(h(x)\geq\delta/\sqrt{n}\) for some positive constant \(\delta\);
Hypothesis~\ref{hyp-dist-bis} then follows from Hypothesis~\ref{hyp-dist} since
\(h(x)/\sqrt{n}\geq\delta/n\geq\delta/\gamma\cdot d(x,f_n(x))\).}
Hypothesis~\ref{hyp-dist-bis}.

\begin{enumerate}
\setcounter{enumi}{2}

\item
\label{hyp-fat}
The faces of \(P_n\) are fat.

\item
\label{hyp-dist}
For any \(x\) on \(P_n\), \(d(x,f_n(x)) < \frac{\gamma}{n}\),
where \(\gamma\) is some positive constant.

\end{enumerate}

Hypothesis~\ref{hyp-fat} is quite natural.
Hypothesis~\ref{hyp-dist} ensures that \(P_n\) approximates \(S\).
Furthermore, the
\(\frac{1}{n}\) factor is reasonable;
indeed, in 2D, when considering a regular polygon with edge length
$\Theta(\frac{1}{\sqrt{n}})$
inscribed in a circle of radius \(1\), the maximal distance between a
point on
the polygon and the circle is \(\Theta(\frac{1}{n})\).
%Indeed, in 2D, when considering a regular
%polygon of complexity \(\sqrt{n}\) inscribed in a circle of radius \(1\), the
%edges have length \(\Theta(1/\sqrt{n})\) and the maximal distance between
%a point on the polygon and the circle is \(\Theta(1/n)\).
The situation
is the same in 3D. Basically it means that the error when approximating
the
surface with a plane is of the second order.

\begin{figure}
\centering
\subfigure[]{
\includegraphics[height=5cm]{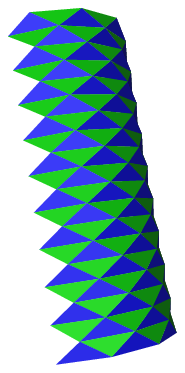}%
\quad
\includegraphics[height=5cm]{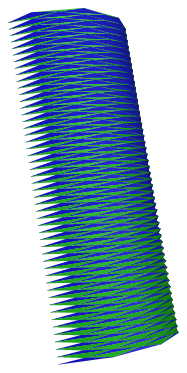}%
\label{fig=schw}
}
\quad \quad \quad
\subfigure[]{
\includegraphics[height=5cm]{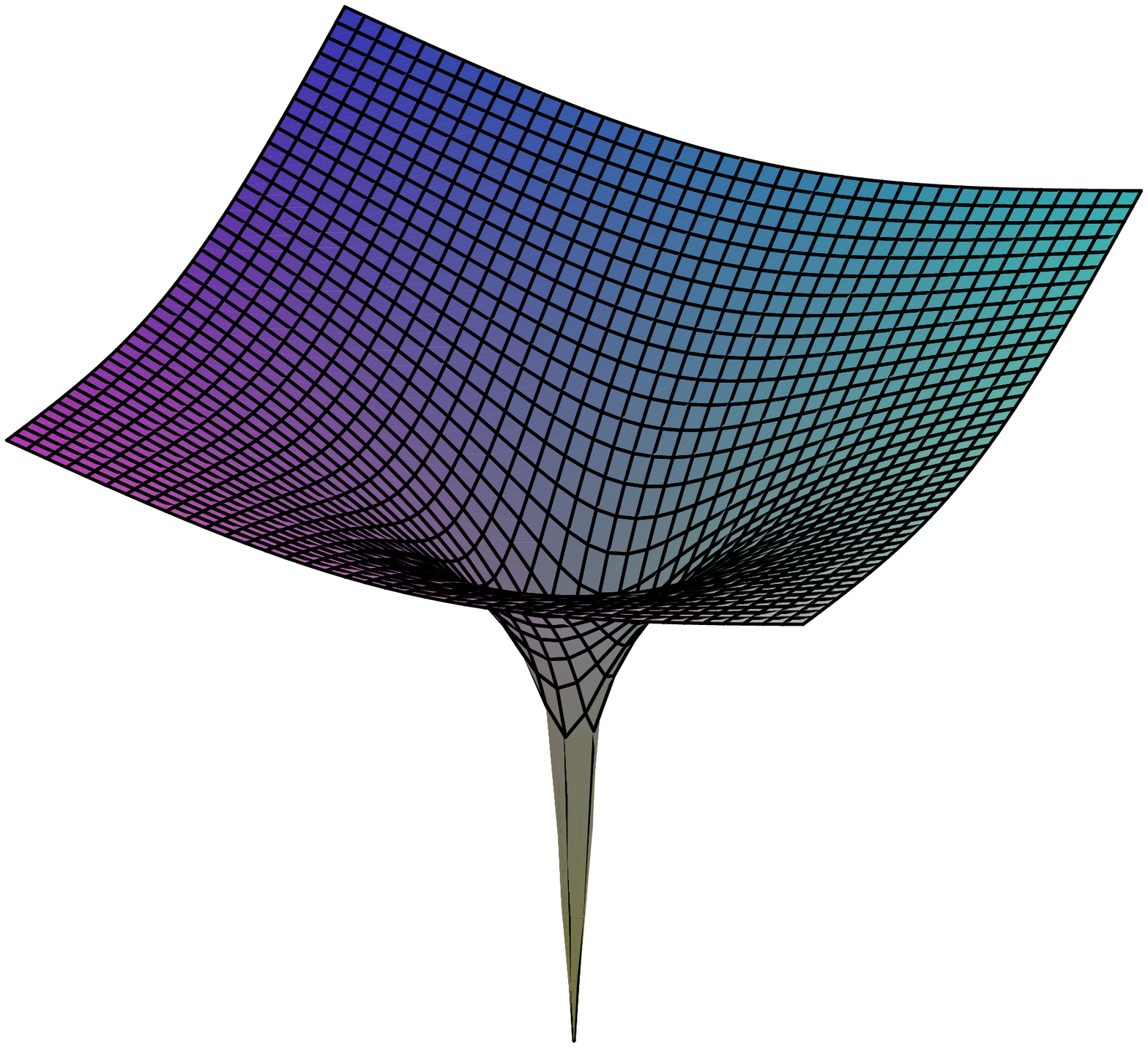}
\label{fig=maple}
}
\caption{(a) Two half lanterns of Schwarz (courtesy of Boris Thibert). (b) A surface that cannot be approximated with the right properties.}
\end{figure}

Our hypotheses (\ref{hyp-length}-\ref{hyp-fat}-\ref{hyp-dist} or
\ref{hyp-length}-\ref{hyp-dist-bis})
%(\ref{hyp-length}, \ref{hyp-fat}, and \ref{hyp-dist} or 
%\ref{hyp-length} and \ref{hyp-dist-bis}) 
ensure that the homeomorphism
\(f_n\) has good properties, that is that, roughly speaking, the
polyhedron can be obtained by only a small perturbation of the surface
while keeping the normal vectors in approximately the same directions.
This is crucial for our proof since otherwise, for example, 
a cylinder can be approximated by a
%``Lampion de Schwarz''
%TODO!!
lantern of Schwarz~\cite{schw} (see Figure~\ref{fig=schw})
whose silhouette has expected complexity \(\Theta(n)\) and unbounded
length.

Notice that the existence of polyhedra with arbitrarily large number of
edges that approximate the
surface according to these hypotheses is a constraint on the surface. Not
every surface admits such an approximation (think of the neighborhood of
\(0\) in the surface defined by \(z=(x^2+y^2)^{1/8}\) as shown in
Figure~\ref{fig=maple}). However, the class of surfaces for which such
approximations exist is quite large. It includes, in particular, smooth
surfaces and polyhedra with fat faces.

%Hypothesis \ref{hyp-captain} is crucial for the whole proof.

\subsection{Proof of Theorem~\ref{th=principal}}
\label{sec=proof}
%\begin{figure}
%\centering
%\includegraphics{alphae}
%\caption{length and dihedral angle of an edge}
%\label{fig=alphae}
%\end{figure}
\begin{figure}
\centering
\subfigure[]{%
\includegraphics{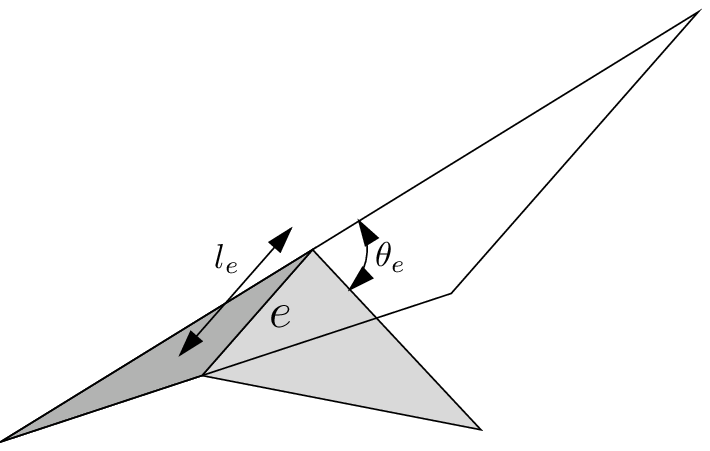}%
}
\subfigure[]{%
\includegraphics{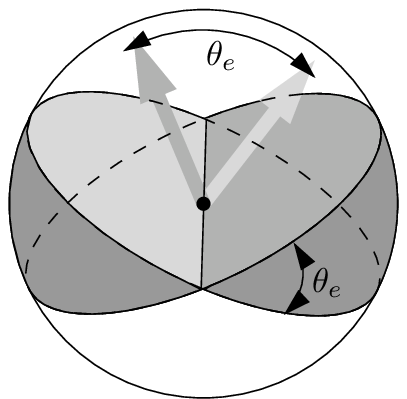}%
\label{fig=alphaeb}
}
\caption{(a) Length and dihedral angle of an edge; (b) set of directions for which $e$ is on the silhouette.}
\label{fig=alphae}
\end{figure}

We consider a point of view chosen randomly at infinity.
We call \(l_e\) the length of an edge \(e\) of polyhedron \(P_n\)
and \(\theta_e\) the exterior dihedral angle associated to \(e\) (see 
Figure~\ref{fig=alphae}).

Let \(T_e\) denote the union of the two triangles adjacent
to edge \(e\) (including \(e\) but not the other edges). For any part
\(\mathcal{R}\) of \(S\), let
\(\mbox{silh}(\mathcal{R})\) be the average length of the part of the silhouette
of \(S\) that lies in \(\mathcal{R}\).

We first recall a classical formula on the expected size of silhouettes
which can also be found, for instance, in~\cite{m-oss-04}.

An edge
\(e\) is on the silhouette if the direction of view is in the dark
area of the sphere of directions of Figure~\ref{fig=alphaeb}. The angular 
measure of this region is \(4\theta_e\), which means that the 
probability
for \(e\) to be on the silhouette is \(\theta_e/\pi\). The expected
number of edges on the silhouette is thus
\[
E=\frac{1}{\pi}\sum_{\texttt{edge }e} \theta_e. \]

We now state our main lemma. The general idea of this lemma is that under
strong hypotheses
(\(S\) has bounded curvature,
the edges have length \(\Theta(\frac{1}{\sqrt{n}})\), and
Hypotheses~\ref{hyp-fat} and \ref{hyp-dist} are satisfied),
one can
prove that \( \theta_e \leq \frac{C}{\sqrt{n}} \) for some constant $C$.
In cases where this inequality does not hold, edge $e$ is near some 
kind of
edge of the surface, or at least some feature that will appear quite
often on the silhouette and we are going to charge this edge to the
silhouette of \(S\).

\begin{lemma}
  For any edge $e$ on $P_n$,
\[ \theta_e \leq \frac{C}{\sqrt{n}} + \frac{8\pi}{l_e}\, \mbox{silh}\left(f_n\left({T_e}\right)\right)
\quad\mbox{with}\quad C=31.3\,\beta.%41.5\,\beta.
\]
\label{mainlemma}
\end{lemma}

Theorem~\ref{th=principal} follows from Lemma~\ref{mainlemma}. Indeed,
since $P_n$ has \(\frac{3n}{2}\) edges, each of length at least $ 
\frac{\alpha}{\sqrt{n}}$ (by
Hypothesis~1), we get that the expected complexity of the silhouette is
%\[E \leq \frac{3C}{2\pi}\, \sqrt{n} +
%8\,\frac{\sqrt{n}}{\alpha}\,  3\ \mbox{silh}(S),\]
\[E \leq \frac{1}{\pi}\frac{3n}{2}\frac{C}{\sqrt{n}}
+ 8\,\frac{\sqrt{n}}{\alpha}\,  3\ \mbox{silh}(S),\]
 because $\displaystyle\sum_{\texttt{edge }e}
\mbox{silh}\left(f_n\left({T_e}\right)\right) = 3\  \mbox{silh}(S)$ since the
length of the silhouette of $S$ that lies in the image (through $f_n$) of a
triangle is counted three times (once per edge). Hence,
\[E\leq \left(15\,\beta+\frac{24}{\alpha}\,\mbox{silh}(S)\right)\, \sqrt{n} = O\left(\sqrt{n} \right).\]
%\[E\leq \left(20\,\beta+\frac{24}{\alpha}\,\mbox{silh}(S)\right)\, \sqrt{n} = O\left(\sqrt{n} \right).\]

\begin{figure}
\centering
\subfigure[]{%
\includegraphics[height=3.1cm]{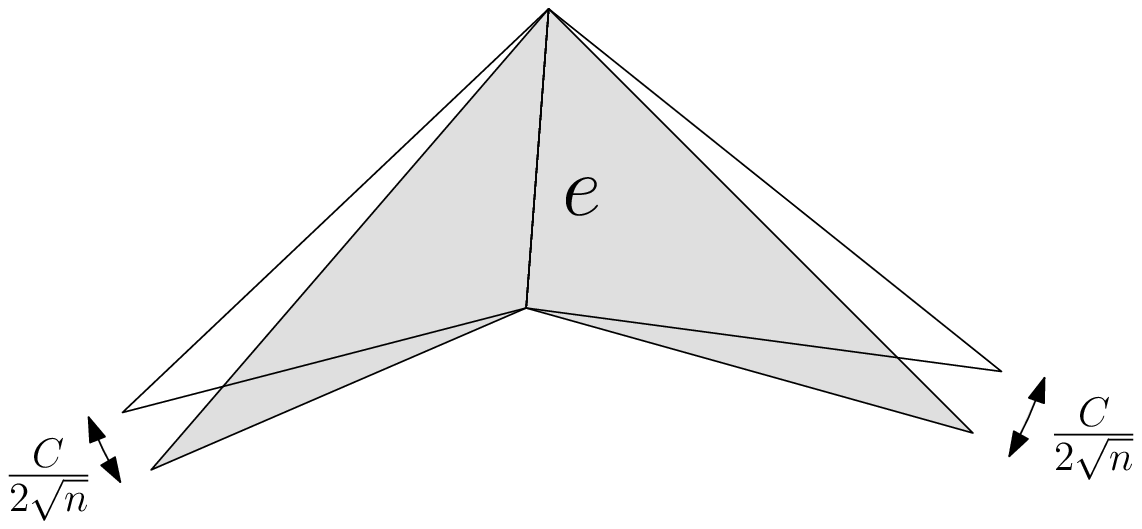}%
\label{fig=rot}%
}
~ ~ ~
\subfigure[]{%
\includegraphics[height=5.2cm]{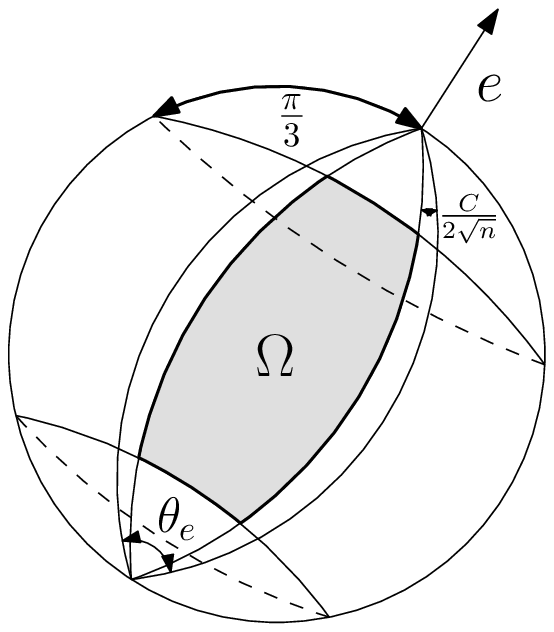}
\label{fig=omega}%
}
\caption{Construction of \(\Omega\).}
\label{fig=rotomega}
\end{figure}

%\noindent{\bf Proof of Lemma~\ref{mainlemma}.}
\begin{proof}[Proof of Lemma~\ref{mainlemma}]
The idea of the proof is as follows.
Consider the set of directions for which \(e\) is on the silhouette. We
first construct a subset \(\Omega\) of these directions whose measure is a
constant times \(\theta_e-\frac{C}{\sqrt{n}}\) (see
Figure~\ref{fig=rotomega}). We then prove a lower bound on the length of the  silhouette
of \(f_n(T_e)\) for all these directions, and deduce the result.

Let $C$ be a positive constant, whose value will be defined later
(see Equation~\ref{constraint}).
%; we will  larger than  \(16 \alpha\)
For any edge $e$ on $P_n$, we can assume that \( \theta_e - \frac{C} 
{\sqrt{n}} >0\) since, otherwise,
$\theta_e\leq \frac{C}{\sqrt{n}}$ and there is nothing else to prove.

%The set of directions for which \(e\) is on the silhouette is
% the set of directions between the planes defined by the faces
%adjacent to \(e\). Rotate each face about \(e\) by an angle of
%\(\frac{C}{2\sqrt{n}}\) so that the exterior dihedral angle decreases by
%\(\frac{C}{\sqrt{n}}\). \(\Omega\) is the set of directions between  these
%two new planes that make an angle larger than \(\pi/3\) with the line
%supporting \(e\). The measure of \(\Omega\) is then
%\(2\cdot( \theta_e - \frac{C}{\sqrt{n}} )\), or $0$.

%TODO: preciser dans la figure: *half* of omega
% verifier que le reste est cohérent à ce niveau.

The set of directions for which \(e\) is on the silhouette is
the set of directions between the planes defined by the faces
adjacent to \(e\). Rotate each face about \(e\) by an angle of
\(\frac{C}{2\sqrt{n}}\) so that the exterior dihedral angle decreases by
\(\frac{C}{\sqrt{n}}\) (see Figure~\ref{fig=rot}).
\(\Omega\) is defined to be the set of directions between 
these
two new planes that make an angle larger than \(\pi/3\) with the line
supporting \(e\);
Figure~\ref{fig=omega} shows one component of $\Omega$, the other one
consists of the symmetric set of opposite directions.
The measure of the set of directions between these
two planes is \(4\, ( \theta_e - \frac{C}{\sqrt{n}} )\). Restricting 
this
set of directions to those that make an angle larger than \(\pi/3\) 
with the line
supporting \(e\), we get, by integrating on the sphere of directions, 
that the
measure of \(\Omega\) is
\(2\, ( \theta_e - \frac{C}{\sqrt{n}} )\).

%If   the measure of \(\Omega\) is  zero, then $ \theta_e - \frac{C} {\sqrt{n}}
%\leq 0$, which implies the result. Assume now that the measure of \ (\Omega\) is
%\(2\cdot( \theta_e - \frac{C}{\sqrt{n}} )>0\).

The remaining step uses the property, which we prove in
Corollary~\ref{cor=techlemma}, that for all the directions in \(\Omega\), the
silhouette of \(f_n(T_e)\) has length at least \(l_e/4\).
Assuming this temporarily, we sum this inequality
over \(\Omega\). The smaller side of the inequality is \(2\frac{l_e}{4}
( \theta_e - \frac{C}{\sqrt{n}} )\). The larger side is the integral of
the length of the silhouette of \(f_n(T_e)\) over all directions in
\(\Omega\), which is smaller than this same integral over all directions,
that is \(4\pi\,\mbox{silh}(f_n(T_e))\). Hence \(4\pi\,\mbox{silh}(f_n(T_e))\geq 
\frac{l_e}{2} ( \theta_e - \frac{C}{\sqrt {n}} )\), which concludes the
proof.
\end{proof}

We now state a lemma and its corollary which we used in the proof of
Lemma~\ref{mainlemma} under the hypothesis that \( \theta_e - \frac{C} 
{\sqrt{n}} >0\). We can thus assume in the sequel that this property holds.

\begin{figure}
\centering
\subfigure[Orthogonal projection along \(d\) of \(e'\) and of the  silhouette of
$f_n(T_e)$.]{%2.6,4,3.5
\includegraphics[height=2.6cm]{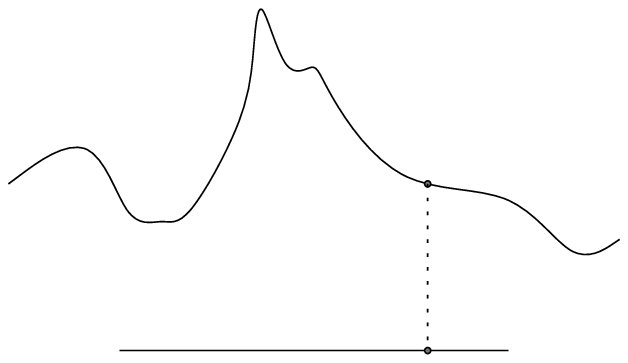}%
\label{fig=projsil}
}
\quad \quad \quad
\subfigure[For the definition of \(D_t\).]{%
\includegraphics[height=3.5cm]{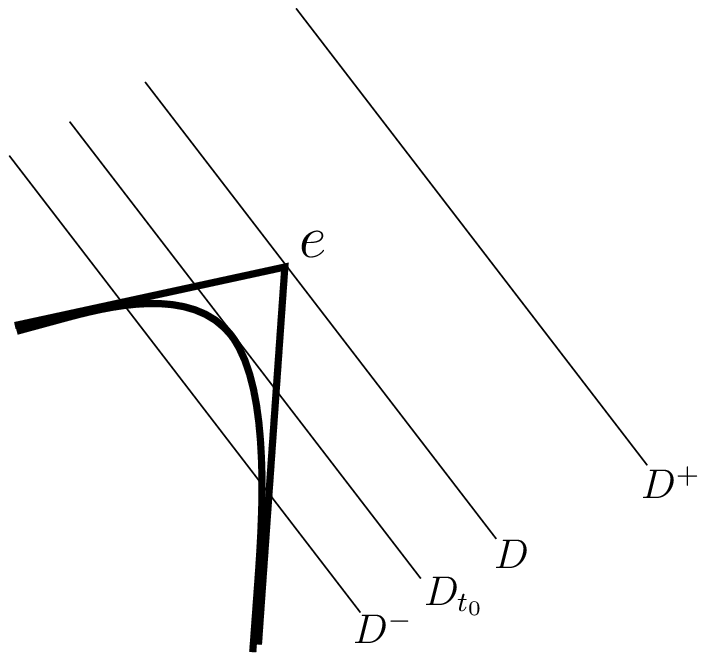}%
\label{fig=sweep}
%\caption{translation of \(D\).}
}
\quad \quad \quad
\subfigure[]{%
\includegraphics[height=2.5cm]{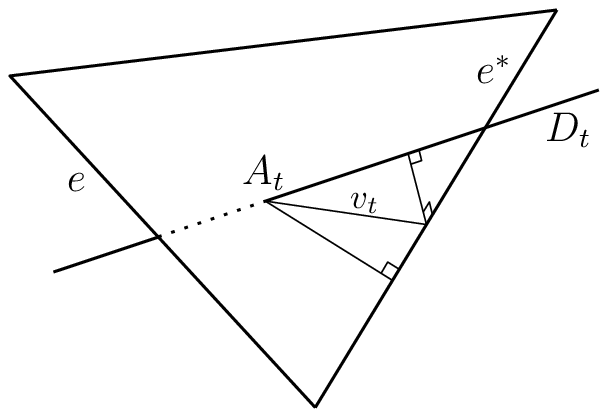}%
\label{fig=uvw}
}
\caption{For the proofs of Lemma~\ref{techlemma} and
Corollary~\ref{cor=techlemma}.}
\label{fig=lemmaa}
\end{figure}

Let $e'$ be the segment obtained by clipping from $e$ all the points at distance
less than $\frac{l_e}{4}$ from its extremities.
Refer now to Figures~\ref{fig=projsil}--\subref{fig=sweep}.
%Refer now to Figure~\ref{fig=lemmaa}(a-b).

\begin{lemma}
  Any line with direction \(d\in\Omega\) that intersects \(e'\)
  can be translated  in a direction orthogonal to
  \(e\) and \(d\) until it becomes tangent to \(S\) in \(f_n(T_e)\).
\label{techlemma}
\end{lemma}

\begin{corollary}
For any direction \(d\) in \(\Omega\), the silhouette of \(f_n(T_e)\) has
length at least \(\frac{l_e}{4}\).
\label{cor=techlemma}
\end{corollary}
\begin{proof}
Consider the projection of $e'$ and of the silhouette of \(f_n(T_e)\) onto a
plane orthogonal to $d$
(see Figure~\ref{fig=projsil})%
. It follows from Lemma~\ref{techlemma} that, in
that plane, each point on the projection of $e'$ maps to a point on the
projected silhouette in the direction orthogonal to $e'$%
. Hence, the projected silhouette is longer than the
projection of $e'$, which is at least $\frac{\sqrt{3}}{2}$ times the  length of $e'$ since $d$
makes an angle of at least $\pi/3$ with $e'$. Thus the silhouette of
\(f_n(T_e)\) has length at least \(\frac{\sqrt{3}}{2} \frac{l_e}{2}>\frac{l_e}{4}\).
\end{proof}

\begin{proof}[Proof of Lemma~\ref{techlemma}]
Let $D$ denote a line with direction \(d\in\Omega\) that intersects \(e'\).
Let $T_1$ and $T_2$ denote the two triangles adjacent to $e$ and let $h_1$ and
$h_2$ denote their respective smallest heights. Let 
$\chi_i = {\beta}h_i /\sqrt{n}$, $\chi^+=\max(\chi_1,\chi_2)$, and $\chi^-=\min(\chi_1,\chi_2)$.
Refer now to Figure~\ref{fig=sweep}. We call $D_t$, 
$t\in[-\chi^-,\chi^ + ]$, 
the line obtained by translating $D$ at
distance $|t|$ in a direction orthogonal to the plane defined by \(e\) and \(d\);
positive values of $t$ correspond to lines in the half-space bounded by the
plane defined by $e$ and $D$, and not containing $T_e$;
%(see Figure~\ref{fig=sweep}). 
negative values of $t$ correspond to lines in the other half-space.
For clarity, we denote 
$D_{-\chi^-}$ by $D^-$ and $D_{\chi^+}$ by $D^+$.

By construction, $D^+$ is at distance $\chi^+$ from $T_e$.  Thus $D^+ $ does not
intersect \(f_n(T_e)\), by Hypothesis~2.
We prove that $D^-$ intersects \(f_n(T_e)\) and that no line
$D_t$ intersects the boundary of \(f_n(T_e)\). This will imply that, sweeping $D_t$
from \(D^+\) to \(D^-\), the first line \(D_{t_0}\) that intersects \(f_n(T_e)\)
is tangent to \(f_n(T_e)\) at one of its interior point, which will conclude the
proof.

We first prove that no line $D_t$ intersects the boundary of \( f_n (T_e)\).  In
other words, we prove that, for each edge $e^*$ on the boundary of $T_e$, no
line $D_t$ intersects $f_n(e^*)$. Let $T_i$ be the triangle (of $T_e$)
containing $e^*$. By Hypothesis~2, it is sufficient to prove that the distance
between $D_t$ and $e^*$ remains greater than or equal to $\chi_i$ for all $t$.

First notice that it is sufficient to prove that the distance between
$D_t$ and $e^*$ remains greater than or equal to  $\chi_i$ for all $t\in[-\chi^-,0]$.
Indeed, then, the distance between $D_0=D$ and $e^*$ is at least
%TODO: qu'en sait-on ?
$\chi_i$, and the distance between $D_t$ and $e^*$ increases for $t\geq
0$ (see Figure~\ref{fig=sweep}).

Let \(\Gamma\) be the smallest angle \(d\) can make with the plane containing
\(T_i\)
%The definition of $\Omega$ yields that \(\tan\Gamma = 
%\frac{\sqrt{3}}{2} \sin\frac{C}{2\sqrt{n}}\).  
and refer  to Figure~\ref{fig=uvw}. Let $A_t$ be the point of intersection between $D_t$ and the
plane containing \(T_i \) and $v_t$ be the distance between $A_t$ and the point
on $e^*$ that  realizes the distance between $D_t$ and $e^*$. The distance
between \(D_t\) and \(e^*\) satisfies  $d(D_t,e^*)\geq v_t\sin\Gamma \geq
d(A_t,e^*) \sin\Gamma$. Hence,  for proving that $d(D_t,e^*)\geq\chi_i$ for  $t \leq
0$, it is sufficient to prove that $d(A_t,e^*)\geq
\frac{\chi_i}{\sin \Gamma}$ for $t\leq 0$. We set $a= \frac{\chi_i}{\sin \Gamma }$ to simplify the
notation. 

We just proved that $d(A_t,e^*)\geq a$ implies $d(D_t,e^*)\geq\chi_i$ (for all
$t$). Conversely, we have that $d(D_t,e^*)<\chi_i$ implies $d(A_t,e^*)<a$.
Similarly, for edge $e$, we get that $d(D_t,e)<\chi_i$ implies $d(A_t,e)< a$. By
definition of $D_t$, we have that $d(D_t,e)<\chi_i$ for $t\leq 0$, thus
$d(A_t,e)< a$ for $t\leq 0$.  
%Now, the definition of \(\Omega\) and the
%direction in which \(D_t\) is translating imply that \(A_t\) moves along a
%straight line in \(T_i\) that makes an angle at least \(\pi/3\) with \(e\).
Furthermore, the angle between $e$ and segment $\{A_t\mid
t\in[-\chi^-,\chi^+]\}$ is at least $\pi/3$ because this angle is at least the angle
between their orthogonal projection on the plane defined by $e$ and $D$ that is
the angle between $e$ and  $D$ since all $A_t$ lie in the plane spanned by $D_t$
which projects on $D$; the lower bound of $\pi/3$ follows since  the angle between $e$ and $D$ is at least
$\pi/3$ by definition of $\Omega$.  
Hence, the locus of points $A_t$, for $t\leq 0$, lies in a region, denoted
$\Upsilon$, shown in dark gray in Figure~\ref{fig=2d}. For proving that
$d(A_t,e^*)\geq a$ for $t\leq 0$, it is thus sufficient to prove that this
region does not intersect the set, denoted $\Upsilon'$, of points %in $T_i$ 
at distance less than $a$ from $e^*$ 
%TODO: ? in T_i, est-ce juste ?
%the two edges of $T_i$ distinct from $e$ 
(shown in light gray in Figure~\ref{fig=2d}).
%
%$\{A_t\mid
%t\in[-\chi^-,\chi^+]\}$ is a segment in $T_i$ that makes with $e$ an angle
%greater than or equal to $\pi/3$ 

\begin{figure*}
\centering
\subfigure[]{%
\includegraphics[height=4.5cm]{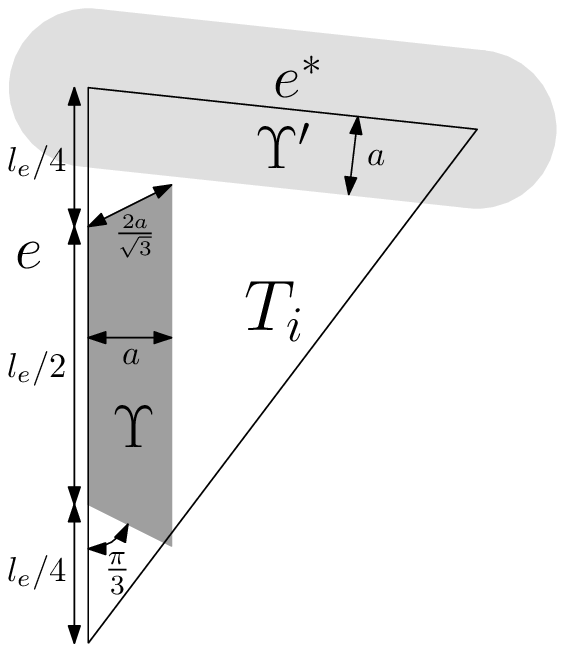}%
\label{fig=2d}
%\caption{free zone.}
}
\quad \quad \quad
\subfigure[]{%
\includegraphics[height=4cm]{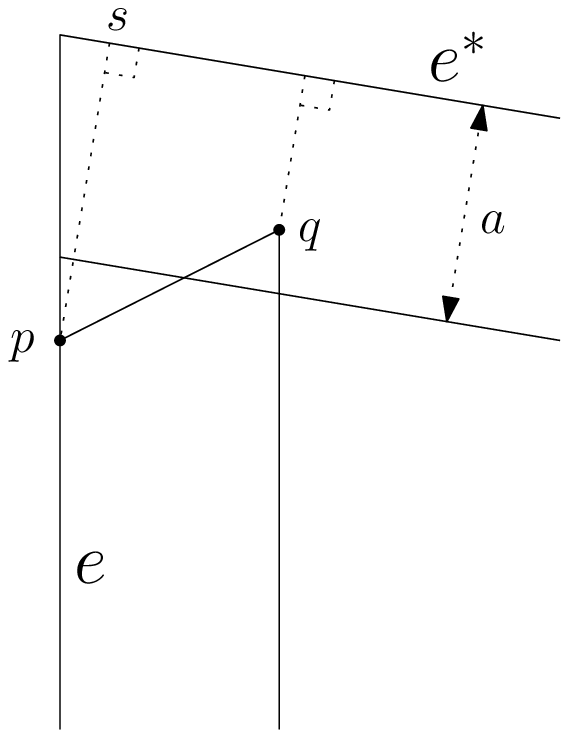}%
\label{fig=2dzoom}
}
\quad \quad \quad
\subfigure[]{%
\includegraphics[height=4cm]{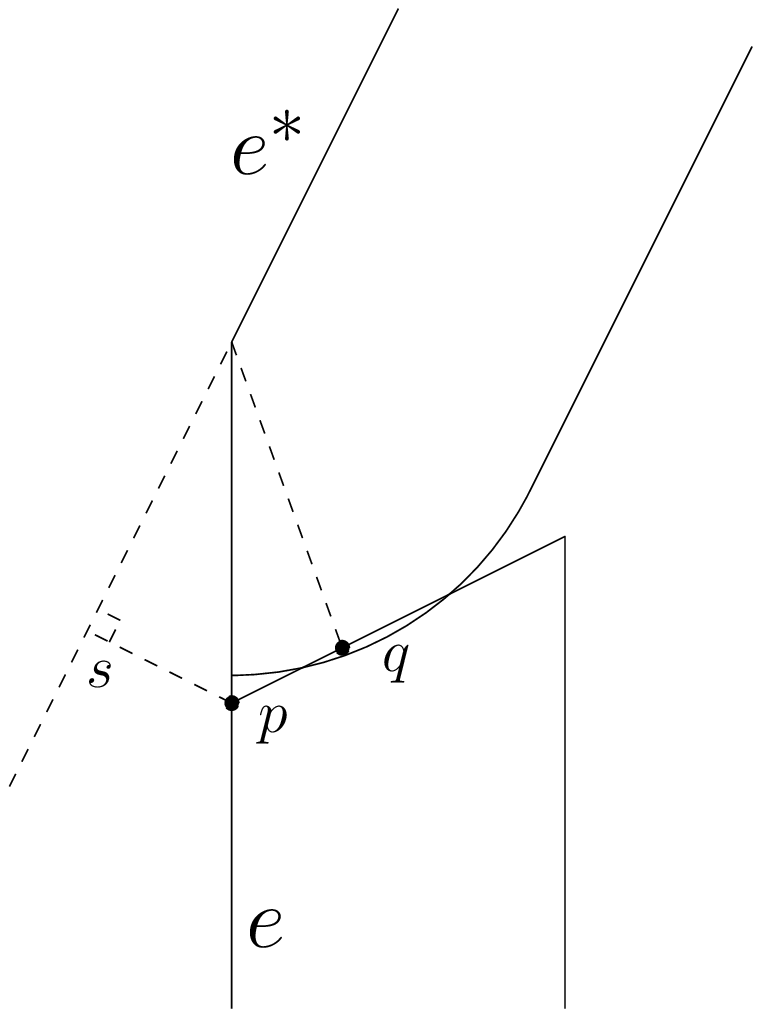}%
\label{fig=2dtordu}
}
\caption{For the proof of Lemma~\ref{techlemma}.}
\label{fig=lemmab}
\end{figure*}

Referring to
Figures~\ref{fig=2dzoom}--\subref{fig=2dtordu}, let $p$ be the endpoint of $e'$ the closest to
$e^*$ and $s$ be its projection on the line supporting $e^*$. 
If the two regions $\Upsilon$ and
$\Upsilon'$ intersect, 
there exists a point $q$ in the intersection that is at distance less
than or equal to $\frac{2}{\sqrt{3}}a$ from
$p$ and at distance less than or equal to $a$ from $e^*$; thus 
$d(p,s)\leq d(p,e^*)\leq d(p,q)+d(q,e^*)\leq (1+\frac{2}{\sqrt{3}})a$. 
%the distance $\|ps\|$ is at most $(1+\frac{2}{\sqrt{3}})a$. 
On the other
hand, $d(p,s)$ is one fourth
of one of the heights of the triangle $T_i$ and thus is at least $\frac{h_i}{4}$.
Hence, if the two regions intersect, then $\frac{h_i}{4}\leq
\left(1+\frac{2}{\sqrt{3}}\right)\,
\frac{\chi_i}{\sin \Gamma}$.
We postpone to Lemma~\ref{lem=techn} the proof that, with
%$C=41.5\,\beta$,
$C=31.3\,\beta$,
we have  $\frac{h_i}{4}>
\left(1+\frac{2}{\sqrt{3}}\right)\,\frac{\chi_i}{\sin \Gamma}$, which implies  that the two regions $\Upsilon$ and $\Upsilon'$ are disjoint. 
This %proves that $d(D_t,e^*)\geq \chi_i$, for all $t$, and thus  
concludes the proof that    no line $D_t$ intersects the boundary of \( f_n
(T_e)\).

\begin{figure}
\centering
\includegraphics[height=4cm]{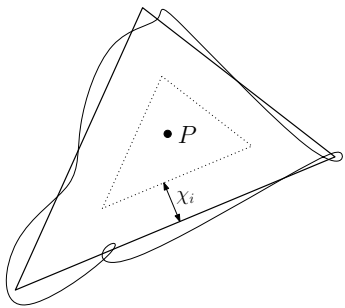}%
\caption{Projection of \(\partial T_i\), \(f_n(\partial T_i)\)
and~\(D^-\).}
\label{fig=triang}
\end{figure}

%%% CESTICI
%%%%%%%%%%%%%%%%%%%%%%%%%%%%%%

We now prove that \(D^-\) intersects \(f_n(T_e)\).  Consider a 
projection, $p()$, along the direction \(d\) onto a plane orthogonal to $d$.  We  proved that, for any
of the two triangles $T_i$, $\Upsilon$ is at distance at least $\chi_i$ from each edge $e^*\neq e$ of \(T_i\). It
follows that $\Upsilon$ lies in triangle $T_i$ and thus that $D_t$ intersects
$T_i$ for all $t\leq 0$. Therefore, $D^-$ intersects $T_i$ and is at distance at
least \(\chi_i\) from each edge $e^*\neq e$ of \(T_i\), for $i=1,2$.
Furthermore, \(D^-\) is at distance
$\chi^-=\min(\chi_1,\chi_2)$ from $e$, by definition. 
We now consider the triangle  \(T_i\) for which $\chi_i=\chi^-$.
It follows that  \(D^-\)
is at distance at least \(\chi_i\) from all three edges of \(T_i\).
Thus \(D^-\) projects to a point $P=p(D^-)$ inside triangle \(p(T_i)\), at distance at least
\(\chi_i\) from the three edges of $p(T_i)$ (see Figure~\ref{fig=triang}).

%Consider now a projection,
%$p()$, along the direction \(d\) onto a plane orthogonal to \(d\) and let
%$P=p(D^-)$.  
%We have
%proved that \(D^-\) is at distance at least \(\chi_i\) from each
%edge $e^*\neq e$ of \(T_i\), for $i=1,2$. Furthermore, 
%
%
%Furthermore, \(D^-\) is at distance
%$\chi^-=\min(\chi_1,\chi_2)$ from $e$, by definition. 
%We now consider the triangle  \(T_i\) for which $\chi_i=\chi^-$.
%It follows that  \(D^-\)
%is at distance at least \(\chi_i\) from all three edges of \(T_i\).
%Thus \(D^-\) projects to a point $P=p(D^-)$ inside triangle \(p(T_i)\), at distance at least
%\(\chi_i\) from the edges (see Figure~\ref{fig=triang}).  
%

%%%%%%%%%%%%%%%%%%%%%%%%%%%%%%

%We now prove that \(D^-\) intersects \(f_n(T_e)\).  
%Consider a 
%projection, $p()$, along the direction \(d\) onto a plane orthogonal to $d$.  
%We have proved that \(D^-\) is at distance at least \(\chi_i\) from each 
%edge $e^*\neq e$  of \(T_i\), for $i=1,2$. Furthermore, \(D^-\) is at distance
%$\chi^-=\min(\chi_1,\chi_2)$ from $e$, by definition. 
%We now consider the triangle  \(T_i\) for which $\chi_i=\chi^-$.
%It follows that  \(D^-\)
%is at distance at least \(\chi_i\) from all three edges of \(T_i\).
%Thus \(D^-\) projects to a point $P=p(D^-)$ inside triangle \(p(T_i)\), at distance at least
%\(\chi_i\) from the edges (see Figure~\ref{fig=triang}).  

Roughly speaking, by Hypothesis~2, the curve $f_n(\partial T_i)$ is at distance
less than $\chi_i$ from $\partial T_i$ (the boundary of $T_i$) thus its
projection $p(f_n(\partial T_i))$ is at distance less than $\chi_i$ from the
edges of \(p(T_i)\).  It is thus intuitively clear that $p(D^-)$ intersects
$p(f_n(T_i))$, and thus that $D^-$ intersects $f_n(T_i)$ (and thus
\(f_n(T_e)\)).

More formally, consider the
application \(g_n\) from the triangle \(p(T_i)\) to the plane containing it such
that, for any point $x$ in $T_i$, the point $p(x)$ is sent to the point
$g_n(p(x))=p(f_n(x))$.  
%First notice that, by Hypothesis~2,  $d(p(x),p(f_n(x)))<\chi_i$ for all $x\in
%T_i$. 
We first  prove that 
the curves
\(p(\partial T_i)\) and \(g_n(p(\partial T_i))\) are
homotopic in \(\mathbb{R}^2\setminus P\).
Consider the continuous application 
\[\begin{array}{rccl}
F: &\partial T_i \times [0,1] & \longrightarrow & \mathbb{R}^2\\
&(x,\lambda) & \longrightarrow & \lambda\,p(x) +(1-\lambda)\,g_n(p(x))= \lambda\,p(x) +(1-\lambda)\,p(f_n(x)).
\end{array}\]
$F$ is an homotopy between the 
curves
\(p(\partial T_i)\) and \(g_n(p(\partial T_i))\) in \(\mathbb{R}^2\).
We  prove that the image of $F$ does not contain $P$, which yields the result.
%, that is that $d(P, F(x,\lambda))>0$ for all $(x,\lambda)\in \partial T_i
%\times [0,1]$. 
The triangle inequality gives 
\[d(P,F(x,\lambda))  \geq d(P,p(x)) - d(F(x,\lambda), p(x)).\]
We have already proved that point $P$ is at distance at least
\(\chi_i\) from $p(x)$ for all points $x$ in $\partial T_i$. 
On the other hand, the distance between $p(x)$ and $p(f_n(x))$ is larger than or
equal to the distance between $p(x)$ and 
their barycenter $F(x,\lambda)$, for any $\lambda\in[0,1]$. Hence 
\[d(P,F(x,\lambda))  \geq \chi_i - d(p(x),p(f_n(x))).\]
Finally, since $d(p(x),p(f_n(x)))<\chi_i$ for all $x\in T_i$,  by  Hypothesis~2,
we have that, for all $(x,\lambda)$, $d(P, F(x,\lambda))>0$. Hence, the image of
$F$ does not contain point $P$
 and thus the curves
\(p(\partial T_i)\) and \(g_n(p(\partial T_i))\) are
homotopic in \(\mathbb{R}^2\setminus P\).

Now, we can contract \(p(\partial T_i)\) to a point while
remaining in \(p(T_i)\). Composing this with \(g_n\) gives a contraction of
\(g_n(p(\partial T_i))\) in \(g_n(p(T_i))\). On the other hand,  there is no contraction of
\(p(\partial T_i)\) in \(\mathbb{R}^2\setminus P\) (since $P$ is in $p(T_i)$),
thus there is no contraction of its homotopic curve \(g_n(p(\partial T_i))\) in \(\mathbb{R}^2\setminus
P\). Hence, there exists a  curve that   is contractible in
\(g_n(p(T_i))\) but not in  \(\mathbb{R}^2\setminus P\). It follows  that  \(g_n(p( T_i))\)  is not included in \(\mathbb{R}^2\setminus
P\). Hence  \(P\) is in \(g_n(p(T_i)) = p(f_n(T_i))\). Therefore, 
\(D^-\) intersects \(f_n(T_i)\) and thus \(f_n(T_e)\), which concludes the proof.

\end{proof}

We finally prove the two following  simple technical lemmas which
complete the proof of Theorem~\ref{th=principal}.
Recall that \(\Gamma\) is the smallest angle a direction \(d\in\Omega\) can
  make with the plane containing \(T_i\).  \begin{lemma}
\label{lem=techn1}
$\sin\Gamma = \frac{\sqrt{3}}{2}\sin\frac{C}{2\sqrt{n}}.$
\end{lemma}

\begin{figure}
\centering
\includegraphics[height=5cm]{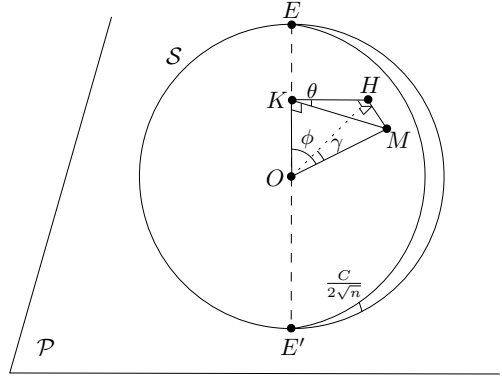}
\caption{For the proof of Lemma~\ref{lem=techn1}}
\label{fig=ouin}
\end{figure}

\begin{proof}
In the following, we identify the sphere of directions with a sphere $\cal S$ embedded in
$\mathbb{R}^3$; let $O$ denote its center. We assume that the embedding preserves
directions ({\it i.e.}, for any direction $d$, the corresponding point $M$ on
$\cal S$ is such that $d$ and $OM$ have the same direction).

Refer to Figure~\ref{fig=ouin}. Let $d$ be a direction in $\Omega$ and $M$ be its
corresponding point on $\cal S$. Consider one of the $T_i$ and  let $\cal P$ be the plane containing $O$ and
parallel to the plane containing  $T_i$.
Let $H$ be the orthogonal projection of $M$ onto plane $\cal P$.
Let $E$ and $E'$ be the two points on $\cal S$ that correspond to the two
(opposite) directions of segment $e$. 
Let $K$ be the orthogonal projection of $M$ (and $H$) onto the line $EE'$.
Finally, let $\theta$ be the angle $\angle MKH$, $\phi$ be the angle $\angle
MOK$, and $\gamma$ be the angle $\angle MOH$. 

It follows from these definitions that \[\sin\gamma =
\frac{HM}{OM}=\frac{HM}{KM}\, \frac{KM}{OM} = \sin\theta\, \sin \phi.\]

Now, the angle $\gamma$ is also the angle between direction $d$ and the  plane that contains
$T_i$. Thus $\Gamma =\displaystyle \inf_{d\in\Omega} \gamma$, by definition of $\Gamma$.
The angle $\theta$ is  the angle between the plane containing $T_i$ and the
plane containing $e$ and $d$. It thus follows from the  definition of $\Omega$
that ${\displaystyle \inf_{d\in\Omega}} \theta = \frac{C}{2\sqrt{n}}$ (see Figure~\ref{fig=rot}). 
The angle $\phi$ is  the angle between $d$ and the line  containing $e$. It thus
also follows from the  definition of $\Omega$
that ${\displaystyle \inf_{d\in\Omega}} \phi = \frac{\pi}{3}$. 
%Furthermore, 
In addition, since $\gamma$, $\theta$ and $\phi$ are in $[0,\frac{\pi}{2}]$, we have
\[\sin\Gamma =\displaystyle \inf_{d\in\Omega} \sin\gamma, \quad \inf_{d\in\Omega} \sin\theta = \sin\frac{C}{2\sqrt{n}} \quad\mbox{and}\quad
 \inf_{d\in\Omega} \sin \phi = \sin\frac{\pi}{3}.\] 
%since $\theta$ and $\phi$ are in $[0,\frac{\pi}{2}]$.
Furthermore, 
the constraints on  $\theta$ and $\phi$ in the definition of $\Omega$ are
independent. Thus, the
minima  of $\theta$ and $\phi$ can be attained for the same direction $d$ in
$\Omega$. 
It follows that 
\[\inf_{d\in\Omega} \left(\sin\theta\,
\sin \phi \right) = \inf_{d\in\Omega} \sin\theta\, . \inf_{d\in\Omega}
\sin \phi. \]
We can thus conclude that 
\[\sin\Gamma =\inf_{d\in\Omega} \sin\gamma = \inf_{d\in\Omega} \sin\theta\,
\sin \phi = \inf_{d\in\Omega} \sin\theta\, \inf_{d\in\Omega}\sin\phi = \frac{\sqrt{3}}{2}
\sin\frac{C}{2\sqrt{n}}.\]
\end{proof}
%\newpage
\begin{lemma}
\label{lem=techn}
$\frac{h_i}{4}> \left(1+\frac{2}{\sqrt{3}}\right)\,\frac{\chi_i}{\sin \Gamma}$
with $C=31.3\,\beta$. %$C=41.5\,\beta$.
\end{lemma}
\begin{proof}
By Lemma~\ref{lem=techn1}, replacing \(\chi_i\) and
$\Gamma$ by their values in the inequality $\frac{h_i}{4}>
\left(1+\frac{2}{\sqrt{3}}\right)\,\frac{\chi_i}{\sin \Gamma}$  gives
\[\frac{h_i}{4}> \left(1+\frac{2}{\sqrt{3}}\right)\,\frac{\frac{\beta
h_i}{\sqrt{n}}}{\frac{\sqrt{3}}{2}\sin\left(\frac{C}{2\sqrt{n}}\right)}\]
or equivalently
\begin{equation}
4\, \beta\, \left(1+\frac{2}{\sqrt{3}}\right) < \sqrt{n}\, \frac{\sqrt{3}}{2}\sin\left(\frac{C}{2\sqrt{n}}\right).
\label{toto}
\end{equation}

Notice first that for large enough
values of $n$, using the approximation \(\sin x\approx
x\) in the neighborhood of zero,
%yields
we derive
the sufficient condition
\[C>\frac{16 \beta}{\sqrt{3}}\left(1+\frac{2}{\sqrt{3}}\right) \sim 19.9\,\beta.\]

Now, since we want our result for all $n$, the computation is more complicated. 
Recall  first that for any strictly  concave
function $f$, such that $f(0)=0$,  
$f(x)> \frac{f(x_0)}{x_0}\, x$ for any $x\in(0,x_0)$. 
It follows that  $\sin x> \frac{2}{\pi}x$ for any
$x\in(0,\frac{\pi}{2})$.
Since we assumed that \( \theta_e - \frac{C} 
{\sqrt{n}} >0\) and thus  that
$0<\frac{C}{2\sqrt{n}}<\frac{\theta_e}{2}<\frac{\pi}{2}$, we get 
\[\sin\left(\frac{C}{2\sqrt{n}}\right)>
 \frac{2}{\pi}\,\frac{C}{2\sqrt{n}}.
\]
To guarantee
inequality~\eqref{toto}, it is thus  sufficient to have
\[4\, \beta\, \left(1+\frac{2}{\sqrt{3}}\right) \leq \sqrt{n}\, \frac{\sqrt{3}}{2}\, \frac{2}{\pi}\,\frac{C}{2\sqrt{n}}.\]
or equivalently 
\[C\geq \frac{8}{3}\left(2+\sqrt{3}\right)\pi\,\beta  \sim 31.27\,\beta,\]
%to guarantee inequality~\eqref{toto}, 
which concludes the proof. 
Note  that we can set 
\begin{equation}
C=31.3\,\beta.
\label{constraint}
\end{equation}
in the definition of $\Omega$ (in the proof of Lemma~\ref{mainlemma}) since Inequality~\eqref{toto} is the only constraint on
$C$.
\end{proof}

\subsection{Generalizations}
\label{sec=general}

We prove here Theorem~\ref{th=principal2}. We first show that
Theorem~\ref{th=principal} generalizes to the case where the viewpoint is
chosen randomly at finite distance. We then show that considering surfaces with
boundary does not change the asymptotic expected complexity of the silhouette.

\paragraph{Point of view at finite distance.}

We have thus far restricted ourselves to the case where the viewpoint is chosen
uniformly at random at infinity. However, our result applies to any distribution
of viewpoints such that the probability for an edge \(e\) to be on the
transparent silhouette is \(O(\theta_e)\), where $\theta_e$ is the exterior
dihedral angle associated to \(e\); indeed, the expected number of edges on the
silhouette is then ${\displaystyle\sum_{\texttt{edge }e}} O(\theta_e)$ and we get
the result by applying, as before, Lemma~\ref{mainlemma}.\footnote{Note that, in
  Lemma~\ref{mainlemma}, $\mbox{silh}(f_n(T_e))$ always refers to an expected 
  length for a viewpoint chosen randomly at infinity.}
Such a distribution of viewpoints is obtained, in particular, when the
point of view is chosen uniformly at random in a ball.
%TODO: in a bounded open region.
This is also the case if \(S\) delimits a bounded region \(\OO\) and the
viewpoint is chosen uniformly at random in \(B\setminus \OO\), for a ball
\(B\).
% that contains \(\OO\) (and such
%that at least a
%constant fraction of the volume of \(B\) is outside of \(\OO\)).

\paragraph{Surfaces with boundary.}

Let  \(S\) be a 2-manifold with boundary \(\mathcal{B}\).
We consider that the boundary is always on the transparent silhouette and
so the definition of the transparent silhouette of
a 2-manifold \(S\) with boundary is exactly
that of a 2-manifold without boundary plus the boundary \(\mathcal{B}\).

The surface \(S\) is approximated by a triangulated mesh \(P_n\) that satisfies
Hypotheses~1 and 2, as in the case without boundary, except that now the mesh
may not be a polyhedron (some edges may have only one adjacent face rather than
two).

To give an upper bound on the number of edges on the silhouette of the
mesh, we consider the boundary edges and the other (non-boundary)
edges separately.
For the non-boundary edges, the same reasoning as before still holds.
For the boundary edges, it is easy to see that the length (in 3D) of the
boundary of \(P_n\) cannot be much larger than the length of
\(\mathcal{B}\). Indeed, the two are homeomorphic, and the hypotheses
imply that the image of an edge \(e\), of length $l_e$, is a curve whose extremities lie at
distance at least \(l_e - 2\beta\cdot \frac{l_e}{\sqrt{n}} = \Omega (
\frac{1}{\sqrt{n}} ) \) apart.
%This means that the number of boundary
%edges of \(P_n\) is bounded by \(O(\sqrt{n})\) times the length of
%\(\mathcal{B}\).
This means that the length of \(\mathcal{B}\) is at least
\(\Omega(\frac{1}{\sqrt{n}})\) times the number of boundary edges of
$P_n$. Hence, the number of boundary edges of $P_n$ is at most
$O(\sqrt{n})$ times the length of \(\mathcal{B}\). So, if the length of
\(\mathcal{B}\) is bounded, the 
expected complexity of the silhouette of
\(P_n\) is \(O(\sqrt{n})\).
%, where the average is taken over all viewpoints at infinity. 
%silhouette of the mesh has complexity \(O(\sqrt{n})\).

\section{Conclusion}
\label{sec=concl}

This paper gives an idea of why, and when, the usual claim that the
silhouette
of a triangulated mesh has size \(O(\sqrt{n})\) is valid.  In
particular, we
have given a set of conditions such that any triangulated mesh
approximating a
surface in a way that satisfies those conditions has a silhouette of
expected
size \(O(\sqrt{n})\).  Roughly speaking, the mesh should have no
short edges,
its faces should be fat, and the distance between it and the surface it
approximates should never be too large.  The surface itself is not
necessarily everywhere differentiable and may have boundaries.

A natural question to ask is whether meshes satisfying those
conditions exist.
In fact, for smooth surfaces, the meshes produced by Boissonnat and
Oudot~\cite{bo-pgsms-05}
are one such example.
%The critical property of
%their  algorithm is that although it produces non-uniform  meshes
%with smaller
%triangles where the curvature is large, the ratio between the size of
%the
%largest and the smallest triangles remains bounded.
The critical property of the meshes
they compute is that the ratio between the size of the largest and the
smallest triangles remains bounded, although meshes are non-uniform with
small triangles in areas of large curvature.
However, in order to satisfy our conditions,  non-smooth surfaces
with curved sharp edges (such as a flying saucer with a sharp
equatorial arc)
would have to be approximated by small
triangles over the whole surface. Such  meshes would have
silhouettes of
expected size \(O(\sqrt{n})\) but then $n$ would be much larger than
necessary; it would be reasonable to replace the large number of
triangles used to mesh large flat portions of the surface with
a smaller number of large triangles%
, which would give a silhouette of size closer to linear. This
%is one of the reasons
explains
why the observed expected size of silhouettes, as shown in
\cite{m-oss-04}, is larger than $O(\sqrt{n})$.
%TODO: et les pieces mecaniques avec tous les sommets sur des "aretes" de
%l'objet et des triangles allonges.
The fact that non-uniform  meshes approximating such surfaces appear,
in computer
graphics, to have  silhouettes of
expected size much smaller than $n$ is thus likely due to additional
properties of the
surfaces or the meshes.
%TODO: j'aime pas cette derni{\`e}re phrase, la virer.

\section*{Acknowledgments}

The authors wish to thank Gert Vegter, who introduced the problem to us, and  gratefully acknowledge fruitful discussions on this topic with
Helmut Alt, Olivier
Devillers, Hazel Everett,   Xavier Goaoc, Bruno Levy and Sylvain Petitjean.

%The authors wish to thank Gert Vegter, who introduced the question,
%Helmut Alt and Xavier Goaoc, who co-authored \cite{agg-owcc-03}, Olivier
%Devillers, Hazel Everett, Bruno Levy and Sylvain Petitjean who helped
%with interesting comments.

%\vfill\eject

%\bibliographystyle{amsalpha}
\bibliographystyle{abbrv}
\bibliography{ploum}

\end{document}